# Farmer-Bot: An Interactive Bot for Farmers

Narayana Darapaneni
*Director - AIML*
*Great Learning/Northwestern University*
*Illinois, USA*
darapaneni@gmail.com

Anwesh Reddy Paduri
*Senior Data Scientist - AIML*
*Great Learning*
*Pune, India*
anwesh@greatlearning.in

Rohit Chaoji
*Student - AIML*
*Great Learning*
*Pune, India*
rohitchaoji@gmail.com

Rajiv Tiwari
*Student - AIML*
*Great Learning*
*Pune, India*
rztiwari@gmail.com

Suman Saurav
*Student - AIML*
*Great Learning*
*Pune, India*
sumansaurav1987@gmail.com

Sohil
*Student - AIML*
*Great Learning*
*Pune, India*
sodhiya2189@gmail.com

*Abstract*— The Indian Agricultural sector generates huge employment accounting for over 54% of country's workforce. Its overall stand in GDP is close to 14%. However, this sector has been plagued by knowledge and infrastructure deficit, especially in the rural sectors. Like other sectors, the Indian Agricultural sector has seen rapid digitization with use of technology and Kisan Call Center (KCC) is one such example. It is a Government of India initiative launched on 21st January 2004 which is a synthesis of two hitherto separate sectors the Information Technology and Agriculture sector. However, studies have shown to have constrains to KCC beneficiaries, especially in light of network congestion and incomplete knowledge of the call center representatives. With the advent of new technologies, like first-generation SMS based and next-generation social media tools like WhatsApp, farmers in India are digitally more connected to the agricultural information services. Previous studies have shown that the KCC dataset can be used as a viable alternative for Chat-bot. We will base our study with the available KCC dataset to build an NLP model by getting the semantic similarity of the queries made by farmers in the past and use it to automatically answer future queries. We will attempt to make a WhatsApp based chat-bot to easily communicate with farmers using RASA as a tool.

## I. Introduction

There is a huge challenge to the Indian agriculture sector, solution to which needs to grow at the faster rate to meet the ever growing demands of the population. This sector, however, lacked much advancement due to lack of technology and investments in Research and Development. However, with invention of new technologies like mobile phones, it became easier to bring this technology to farmers. Many initiatives were started up keeping these new advancements in mind and Kisan Call Center (KCC) was one such initiative started by the government of India. The government made the data of the interaction with farmers publicly available for research and analytic purpose, which is termed as the KCC dataset. The data captures various details which categorizes the query made by the farmer, captures the details of the query, the resolution provided, state, district and the time when the query was raised / resolved. This generates a huge dataset which can be utilized for various research and analytics purposes.

Currently, the KCC is challenged with problems such as network issues, problems in connecting to the centre and lack of trained customer care representatives, which has shown a constrained experience to the KCC beneficiaries. To help mitigate some of these issues, we propose to build a chat-bot which can be trained on this huge corpus to get semantic similarities of historical queries and provide a resolution based on NLP model analytics. The solution will be based on existing medium of communication – WhatsApp, which is common among the masses and channel is via a NLP pertained model to respond the queries. To help us achieve this, we will be using RASA as the tool for building the chat-bot.

Chatbots are made on principle of minimizing the manual work at the call centers, and make human life less demanding in every day prerequisites. Agriculture sector is not new to these changes and has witnessed rapid digitization, including advent of bots for specific purposes. There is a huge range of information that the farmers seek in order to sustain their farming, such as information on weather, rainfall, plant protection, government schemes, information on fertilizers and seeds, etc. This information was historically provided by local government bodies, and later with advent of phone services, could be provided to farmers who contacted them through the centralized call center. Over years this has built up a huge information base which can be utilized by modern analysts to extract relevant information and pass as a resolution of queries to the farmers. The mobile devices and faster means of communication has fostered need to utilize this data and build automated bots to help mass farmers. A few bots which are being used in the agricultural sector currently are:

### A. Farm Chat:

The analysts at FarmChat[6] created an information base for potato cultivation using the KCC dataset, and also collected data from developmental interviews with small holder ranchers and agri-experts. For each of the distinguished subjects, they inquired the two agri-experts and applied illustrations of commonplace farmer questions, the follow-up questions that they would inquire in arrange to get at the issue, and the ultimate resolution they would provide. All such conversations were included to the CSV file, and the informational advice was included within the Farm Chat information base.

## B. AgronomoBot:

For agricutural puposes it is vital that the the information about the field conditions[8], such as soil condition, air, weather, soil moisture, rainfall, windspeed etc be rapidly and easily be available to farmers. This bot was developed based on research of data acquired by wireless sensors placed in a vineyard. The bot was deployed as a tool to the farmers to provide information by integrating the bot to Telegram API. Further advancements are planned for the bot, such as the extension to other informing stages, the usage of discourse communication capacity, picture classification and nonstop information investigation. It is trusted that with explanatory capacity over the mass of accessible information, it helps in early location of infections in crops, vitality and water squander lessening, and progressed administration capabilities for the farmer.

## II. MATERIALS AND METHODS

### A. Data collection

Kisan Call Centre (KCC) data is a set of queries asked by farmers in KCC and the response provided by the KCC. The entire corpus is available on Indian government's official website, "data.gov.in" by the data set name Kisan Call Centre. Data is available from year 2006 to 2020 in CSV format. There is a separate catalogue maintained for each district of each state for every month. Few sample URL for extracting the JSON data for some of the districts in Assam are as below:

http://dackkms.gov.in/Account/API/kKMS_QueryData.aspx?StateCD=02&DistrictCd=0201&Month=01&Year=2015

http://dackkms.gov.in/Account/API/kKMS_QueryData.aspx?StateCD=02&DistrictCd=0202&Month=01&Year=2015

http://dackkms.gov.in/Account/API/kKMS_QueryData.aspx?StateCD=02&DistrictCd=0202&Month=01&Year=2016

http://dackkms.gov.in/Account/API/kKMS_QueryData.aspx?StateCD=02&DistrictCd=0203&Month=01&Year=2015

The dataset consists of the following fields:

- A. Season: Information on which season the question was asked.
- B. Sector: What is the sector about which the question was asked like – Agriculture, Horticulture, Animal Husbandry, Fisheries
- C. Category: Within a sector which category is the question about for example – Vegetables, Animals, Cereals, Drugs and Narcotics etc.
- D. Crop: Which crop the question is about like – Potato, Paddy, Wheat, Banana, Goose etc.
- E. Query Type: Broad category of specific query asked like – Plant Protection, Weather, Nutrient Management, etc.
- F. Query Test: Actual query asked by the user
- G. KCC Answer: Answer provided by the KCC representative to the caller Kisan.
- H. State Name: State in which the question has been asked.
- I. District: District within the state from where the question has been asked.
- J. Block Name: Block of the district where the question has been asked from.
- K. Created On: Date on which the query was asked

Since we wanted to pilot a bot for a particular state, not all columns will be relevant to us. Also, we wanted to pull the latest records, so for this purpose, we have used the last 5 years of records to build the chat-bot.

For the purpose of this research and pilot we have the following fields - Query Type, Query Text and KCC Answer.

We omitted the crops and other categories to generalize the discussion more and to avoid the bot falling into a monotonous state of asking the same details to the user. This would in effect make the communication faster, smoother and less boring for the user. This would also in effect keep the users engrossed making sure they don't lose the interest to use the bot.

| Query Type | Query Text | Query Answer |
|---|---|---|
| Plant Protection | CONTROL OF APHIDS IN PADDY | APPLY INDOFIL M-45/ANTRACOL @ 2 GRAM PER LITRE OF WATER |
| Plant Protection | ASKING CONTROL OF LEAF HOPPER ATTACK IN BORO RICE | SPRAY EKALUX @ 3 ML PER LITRE OF WATER |
| Plant Protection | SEED TREATMENT IN PADDY | TREAT THE SEED WITH MANCOZEB @ 2.5 GM LITER OF WATER FOR 24 HOUR |
| Weather | QUERY REGARDING WEATHER REPORT FOR CACHAR DISTRICT | INFORAMATION GIVEN THAT MODERATE RAIN MAY OCCUR IN COMING 3 DAYS |
| Nutrient Management | Micronutrient FOR CUCUMBER | ADVICED TO SPRAY TRACEL-2 @ 2GM/L OF WATER |

TABLE1.Sample of the data retrieved from the KCC dataset for state of ASSAM.

As can be seen from examples above, the data captured is a summary of discussion with the farmer and would need some manual tweaking to standardize it to normal English format before it can be fed to any NLU processing system. Feeding the crude data will fail miserably on normal human interactions.

The intention of this activity was to create general human interaction corpus so that the same can be used to train the bot. We choose the mode of communication as WhatsApp, which is popular among masses in India and is easily accessible to the users. We harnessed the power of RASA tool to achieve this communication gateway setup.

*B. Statistical Methods*

Natural Language Processing (NLP) is a theory-motivated range of computational techniques for the automatic analysis and representation of human language processing [9]. The NLP research is in a paradigm shift, they are no longer based on techniques of recognition and understanding of loose words. But now begin to explore semantic techniques more consistently, which is a jump from syntactic curve to the semantics curve, and ultimately will arrive at the pragmatics curve, where computational programs will be able to investigate and build entire narratives.

Word2Vec is a shallow 2-layer neural network model that takes a large corpus of text as input and outputs a large vector space with each word corresponding to a unique vector in the space. Models trained with Word2Vec can be trained to understand context of words based on its usage and words surrounding it. This makes the algorithm sensitive to order of words.

Aside from shallow fully connected neural networks, it is also possible to use recurrent neural networks (RNN) for training a natural language model. A recurrent neural network is a fully connected neural network which replaces some of the extra hidden layers with looping of some of its layers. RNN is particularly good at finding patterns in sequential data, such as text in languages that are sensitive to word order. However, RNNs have a problem with long term memory due to vanishing and exploding gradients.

It is possible to add Long Short-Term Memory (LSTM) gates to a recurrent neural network, to improve the long term memory of its weights and also to control how much data is "forgotten" at each step, in order to free up memory for new information (and increase long term retention) [13,14].

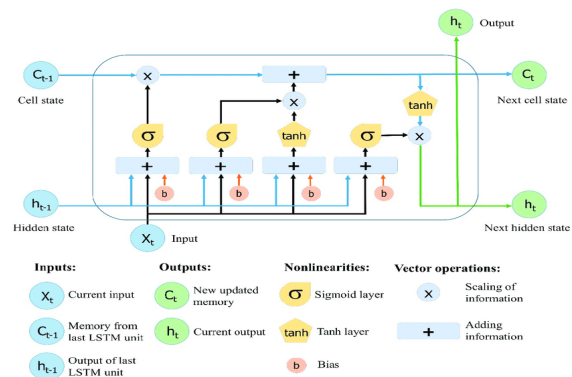

FIGURE 1. Application of Long Short-Term Memory (LSTM) Neural Network for Flood Forecasting [10]

Recent developments in neural networks have given rise to a new technique, called Attention mechanism. The idea is based around focusing "attention" towards only the most important details – in case of language processing, the most important words. This allows us to bypass the problem of long-term retention of memory [12].

In an attention mechanism, the entire text is not vectorized, but instead different parts of a text are read by decoder during output generation. Using this mechanism, the model learns based on input and previous outputs and learnings, and decides the importance of each word. The advantage of attention in NLP is that it is flexible with how it learns sequence of words, and can be used for languages that have a different structure from source Term Frequency-Inverse Document Frequency (TF-IDF) identifies most commonly occurring words in a given text or collection of texts. It is often used to point towards documents which are most relevant to an input query. Term Frequency identifies how often a word occurs in a set of documents, denoting word importance, while Inverse Document Frequency is used to determine the amount of information contained in a word (i.e., the more often it occurs, the less information it is likely to contain, for example "the", "in", "a", etc.). Using TF-IDF and relevant data it is possible to train a model to match new queries with existing ones and return the relevant results.

Term Frequency is mathematically represented as tf(t,d) for term t and document d.

Inverse Document Frequency is represented by idf(t,d) and it is given by formula

$$idf(t, D) = \log \frac{N}{|\{d \in D: t \in d\}|} \qquad (1)$$

Where,
N = Total number of documents in the corpus = |D|
$|\{d \in D: t \in d\}|$ = Total number of documents where the term t appears. Sometimes, it is corrected to 1 + $|\{d \in D: t \in d\}|$ to avoid any zero division errors.

TF-IDF is the product of the term frequency and inverse document frequency and hence is mathematically represented as
tfidf(t,d,D) = tf(t,d) .idf(t,D)

The disadvantage of this approach is that it is heavy on resources and time for longer sentences and larger texts.

Based on the idea of attention, more sophisticated architectures, such as Transformer have been built. Transformer aims to use the principles of Attention mechanism while solving its biggest drawbacks – training time due to large vector sizes, as vectorized sentences can be very long, depending on their length. It uses a method called Self-Attention which allows the model to look at and relate different words in a sentence to each other. The disadvantage of using Transformer is that it can only accept strings with a fixed length as an input, or requires chunking data to be fed into it. This chunking of data can lead to loss of contextual meaning.

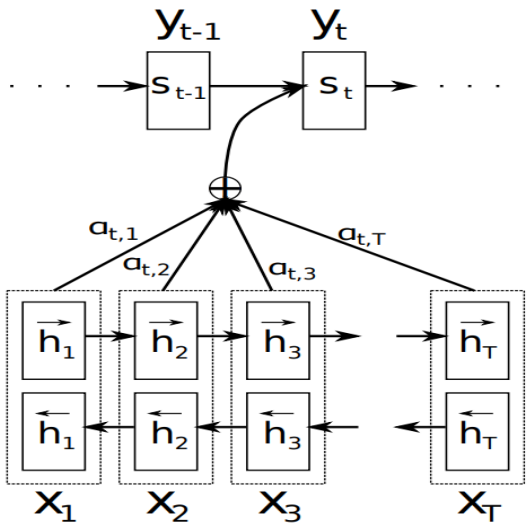

FIGURE 2. Attention Model example

The data being used in our project is Kisan Call Center data. The queries and answers are in form of short, written summaries of voice calls, which were later tweaked to be more natural and conversational. Thus, the disadvantages of certain models that arise from larger sentence lengths are not completely relevant to us, and hence our choice of algorithm can be based on other factors that lead to a balance of accuracy and training time.

C. *Graphical Representation and Exploratory Data Analysis*

For our research and demo purpose, analysis was done on Assam's data set from 2015 to 2020. There are 116163 queries. More than 50% of queries are for Season Rabi, followed by Kharif and Jayad. Sector wise analysis shows that Horticulture and Agriculture almost constitute 85% of the total queries.

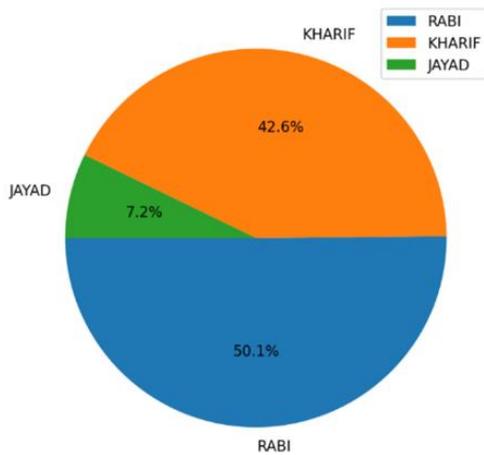

FIGURE3. Crop Types within Agriculture Sector

Analyzing the queries, we observe that they predominantly pertain to the main agricultural seasons in India, namely Kharif and Rabi. This data can be harnessed to decide the infrastructure that would be needed to host the solution and the server requirements can be scaled up or down based on this data.

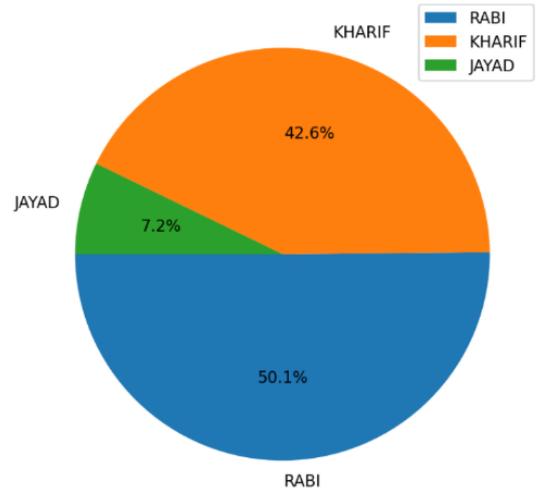

FIGURE 4. Different Sectors in KCC Dataset

The queries by the farmers are distributed across various sectors the most common ones being from the Agriculture and Horticulture. This also coincides with the fact that the primary agriculture sector in Assam is of cereals, vegetables and tea plantations. This distribution can be utilized to make the both more effective.

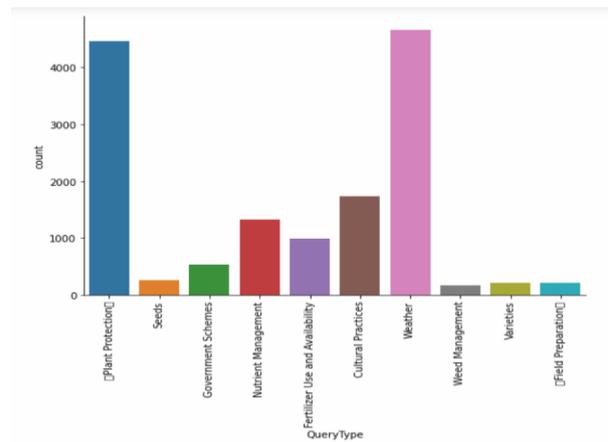

FIGURE 5. Query Types

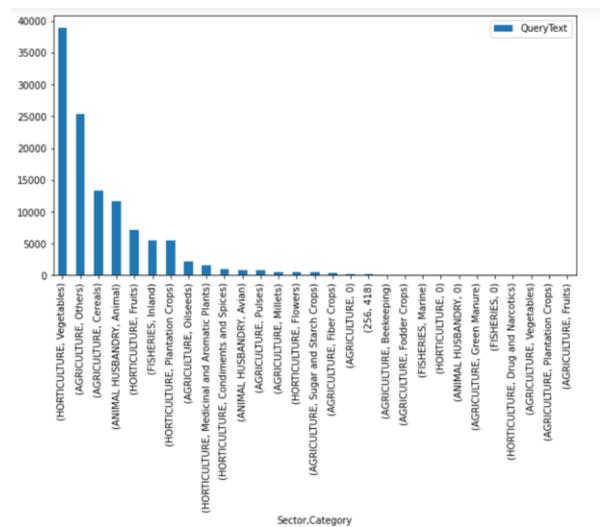

FIGURE 6. Query Text Categories

In the figures above FIGURE 3 and FIGURE 4 we can see the distribution of questions against different categories. We can see that most of the question are in relation to plant protection and weather, and are mostly asked on horticulture and agriculture which is in relation to demographics of Assam.

*D. Tools – RASA and Solution Workflow*

RASA [21] is a tool to build custom AI chatbots using Python and natural language understanding (NLU). It also allows the user to train the model and add custom actions. Chatbots built using RASA can be deployed on multiple platforms like FB messenger, WhatsApp, Microsoft bot and slack etc.

Rasa has two main components:
A. RASA NLU (Natural Language Understanding): RASA NLU is an open-source natural language processing tool for intent classification (decides what the user is asking), extraction of the entity from the bot in the form of structured data and helps the chat-bot understand what user is saying.
B. RASA Core: A chat-bot framework with machine learning-based dialogue management which takes the structured input from the NLU and predicts the next best action using a probabilistic model like LSTM neural network rather than if/else statement. Underneath the hood, it also uses reinforcement learning to improve the prediction of the next best action.

We propose a solution of chat-bot which is intuitive for the farmers and simple to reply their questions in a successful manner The Workflow begins with the hello message answer by the bot. The beneficiaries of the app can then start querying the bot in the topic of their relevance e.g. Agriculture Potatoes. The bot will get the best possible solution based on the prepared dataset, which we deliver while training the bot. The bot will be trained on the KCC dataset.

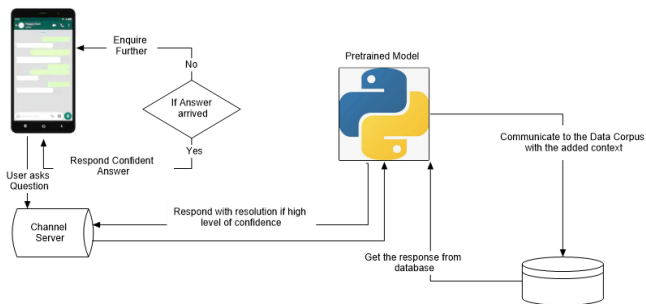

FIGURE 7. High level application overview

The figure above, FIGURE 7, is the high level interaction of the application by the end user. We can see that the data display or user interaction happens via a mobile device which just acts as medium to send and receive messages (via WhatsApp). The data is passed via a channel to a pre-trained model hosted on a server, which finds the correct intent of the user and correspondingly responds to the query with the appropriate response selection. We can deep dive into the response selection feature from RASA via the below FIGURE 8.

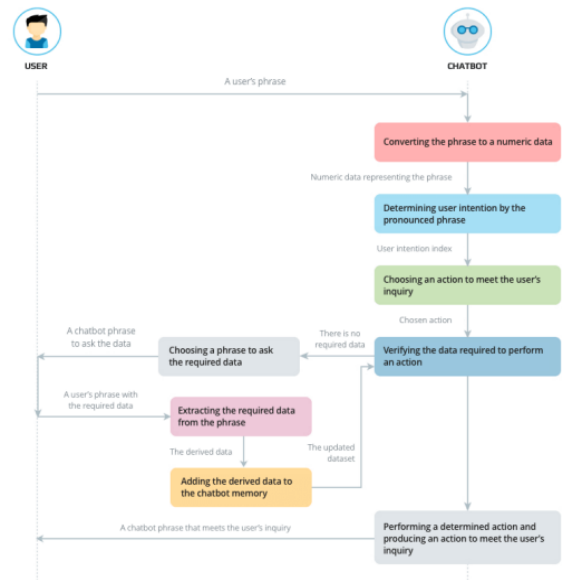

FIGURE 8. RASA contextual data handling.

The details of the data processing and contextual data handling in RASA can be viewed by the diagram above FIGURE 8. As can be seen from the interaction flow, the RASA NLU converts the phrase (query from user) into a numeric value and tries to determine the intent via algorithms provided. In this, we use the TF-IDF to identify the closest phrase match. Based on the match it generates and action, which in turn checks if some data is needed to perform that action. Once the preconditions are met, the RASA NLU will pick the response to the selection action and send it to the end user. This will be processed via channel layer and provided to the end user as chat response on WhatsApp.

III. RESULTS

The Farmer-Bot is designed to seamlessly answer queries of the beneficiaries and is an attempt to provide the mass farmers a communication channel through which they can ask their queries and get resolution at any time without needing to worry about call center timings and network congestion issues. Also since WhatsApp is one of the leading chat mediums in India, with a mass reach we choose to integrate the chat-bot to the WhatsApp platform.

The bot will respond to greeting by the user with an appropriate greeting message of introduction about itself, and then the user can proceed with their queries. After answering the query, the bot will check for the satisfaction of the answer, if satisfied, it will welcome user for any further query. If answer is not satisfactory, the bot will provide the details of the call center to the farmer to contact on phone during the working hours.

The RASA platform on which the bot has been built provides a number of ways to integrate the NLP algorithms to build out a solution. We can use the built in models or feed any custom model to the platform and build the final model via transfer learning using the data provided to it.

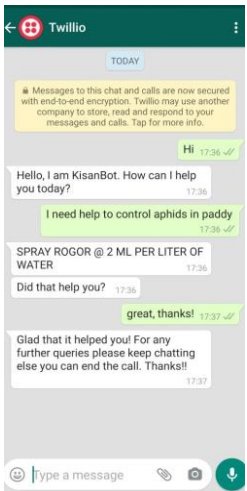
FIGURE 9. Example of Successful Intent and Response

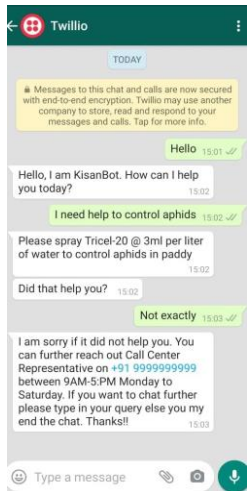
FIGURE 10. Example of user prompted unsuccessful response

The above figures show the response from the bot for scenarios where the confidence rate was high enough to give the answer to the user. We have put a threshold limit of 0.7 or 70% as the confidence level to let the bot send back a response. In FIGURE 10, we can also see an alternate route when the bot might reply with an incorrect answer to a query as per beneficiaries' follow-up response. In such cases, the bot apologizes and provides the user with alternate option, in this case to call the KCC during working hours.

In response to the FAQ (frequently asked questions), a beneficiary may also ask some irrelevant queries, which the bot may not be able to answer. In that case the bot will go for a Fallback response since the confidence level for answering the query will be less than 70%.

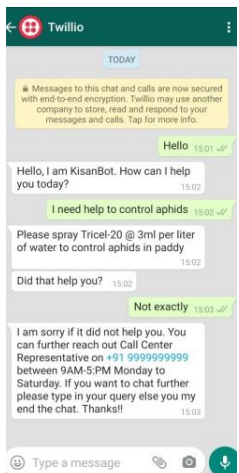
FIGURE 11. Example of fallback

All these predictions are based on the fallback confidence level of 0.7 or 70%. We have tested the built model and found it to be satisfactory in providing a relevant response.

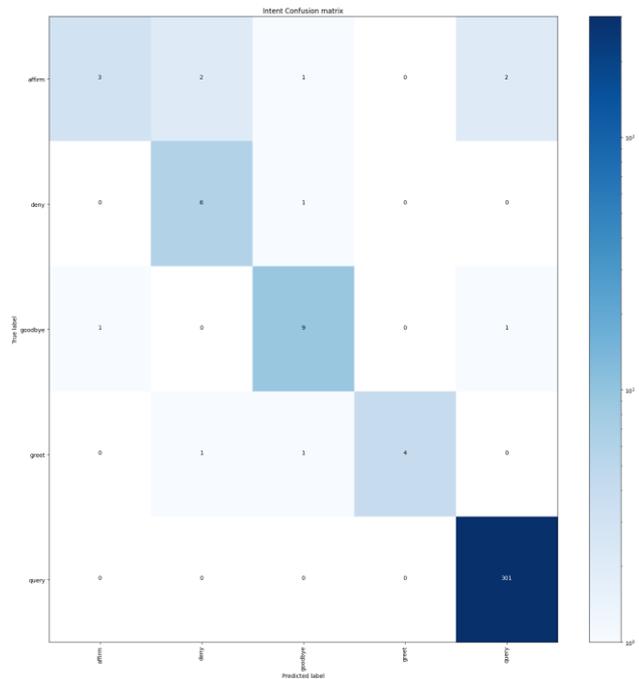
FIGURE 12. Intent Confusion Matrix

As can be seen with the given data set the mode is performing quite well. There are a few miss-interpreted response as evident in the FIGURE 12 confusion matrix above.

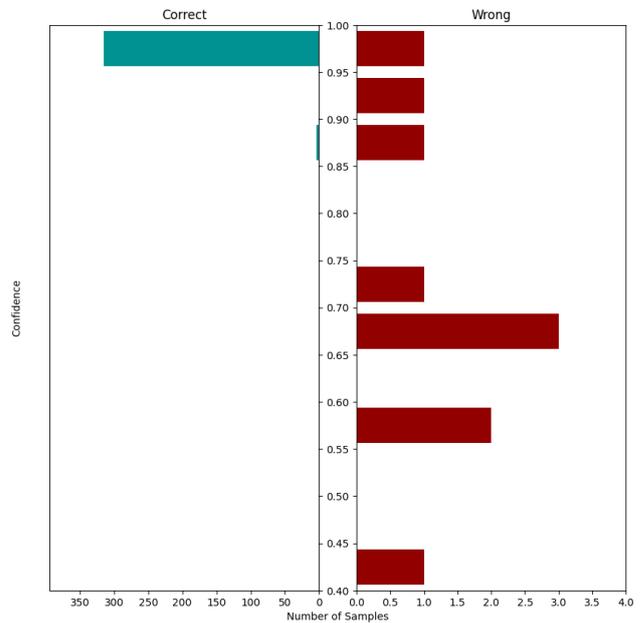
FIGURE 13. Intent Prediction with Confidence Distribution

We can see from the histogram above that there are a very high number of intents correctly identified with above 95% confidence. There are some intents with lower confidence but above threshold. The number of these intents is quite less. There are some intents which are predicted incorrectly, but most of them are in lower confidence range.

## IV. DISCUSSION AND CONCLUSION

### A. Discussion on the results

The Indian agriculture is huge employment sector with over half of the population dependent on this for their living. The mass population needs information of various kinds like weather, seed, crop protection, schemes etc. The KCC was established for this purpose and has been serving the farmers for over couple of decades now. However the lack of enough call centers and poor network connectivity hampers this flow of information critical to the farmers. Farmer-Bot is an attempt to bridge this gap by picking the benefits from either end. On one had we attempt to provide the information to farmers on the tip of their finger via a WhatsApp chat channel and on the other hand utilizes the power of huge KCC dataset.

In our attempt to pilot the bot, we have utilized a small part of Assam's KCC data set for past 5 years to build out an FAQ-like chat bot which can answer based on similar queries asked in the past. Our aim was to get the bot to respond to generalized queries by the beneficiary. In our minified dataset, the bot responds quite well to the queries. As can be seen in the confusion matrix –FIGURE 10 the accuracy of the model to respond to human queries is about 96%

As per our, results we can see that for the given dataset the bot beautifully replies to the similar queries. As per the histogram graph on FIGURE 13, we can see that most of queries which the bot responded correctly are with a very high level of confidence above 95%, there are a few which responds in the range of 85%, whilst there are a very few which have responded wrong answer with high confidence level but majority of answers responded incorrectly were with a low confidence. This signifies that the model is working well with a few false positives.

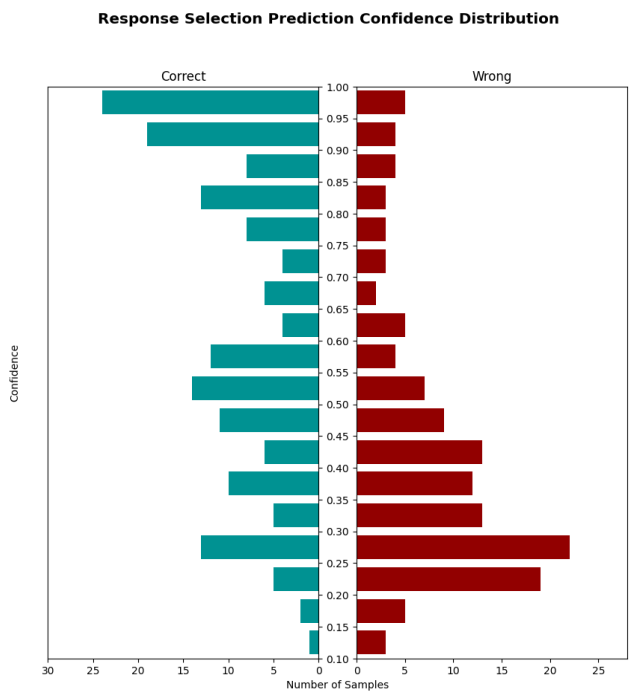

FIGURE 14. Response Selection Prediction Confidence

In the histogram graph above FIGURE 14 we can see the histogram of the plot of correct v/s wrong response selection and we can see that most of the responses selected correctly are at higher confidence while there are a very few response selected with lower confidence level which are correct. It is exactly upside down for the incorrectly selected response cases. There are fewer with higher confidence as compared to lower confidence. This again provided a confidence on the stability of the model.

### B. Limitations in the study

At present, the study is limited to a very small data set, and as we know the KCC has a huge database we need to see that how will this prediction model perform on larger dataset.

The model also appears to be a bit biased at the moment, as the generalized queries are still specific and the model is easily able to catch the keywords. However, we will have to test the model with a bigger dataset and see how it performs on more generalized data.

Besides, we have also used the data set from one state, Assam, where the data was all in English language. There is a huge corpus of data in regional languages and we need to validate the model performance on these regional languages.

### C. Conclusion

Farmer-Bot is one the first attempt to bring two hitherto different business domains together for a noble cause to bring the information quickly and hassle-free to farmers of India. It is an agglomeration of harnessing big data technology and mobile technology to bring a seamless user experience via WhatsApp to the end beneficiaries (farmers) in helping them resolve their queries at any point of time. This will save the cost of call center on one hand and will help the beneficiaries get the information at any point of time without having to wait for KCC to open or worry about network congestion issues.

The solution also brings a balancing act where it will respond to queries which it can interpret from the existing corpus, or else will provide the end users a seamless experience to get the support to call center by providing the call center number.

The solution presented in this paper is just a POC work, but the outcome of the sample testing is encouraging and can definitely be pursued further and built into an enterprise solution. In the next step we can also add solutions to build more corpus based on any new query being asked by the user and its resolution based on solution provide by the KCC representative.

## REFERENCES

[1] Rahul Wagh and Anil Dongre, "Agricultural Sector: Status, Challenges and its Role in Indian Economy," Journal of Commerce and Management Thought, DOI: 10.5958/0976-478X.2016.00014.8.

[2] S. K. GOYAL, Prabha, Dr Jai P. Rai, Shree Ram Singh, "Indian Agriculture and Farmers-Problems and Reforms", In book: Indian Agriculture and Farmers (pp.79-87), January 2016 Researchgate.net. [Online]. Available: https://www.researchgate.net/publication/330683906_Indian_Agriculture_and_Farmers-Problems_and_Reforms

[3] S.Kavitha, Nallusamy Anandaraja, "Kisan Call Center Services to the Farming Community: An Analysis", *Journal of Extension Education,* Vol. 29, No. 3, 2017.

[4] S. Kavitha and N. Anandaraja, "Constraints and Suggestions as Percived by the Kisan Call Center Beneficiaries", *Int. J. Pure App. Biosci. 5 (4): 1725-1729 (2017),* ISSN: 2320 – 7051

[5] Devesh Thakur, Mahesh Chander, Sushil Sinha, "WHATSAPP FOR FARMERS: ENHANCING THE SCOPE AND COVERAGE OF TRADITIONAL AGRICULTURAL EXTENSION",

*International Journal of Science, Environment and Technology,* Vol. 6, No 4, 2017, 2190-2201.

[6] Mohit Jain, Pratyush Kumar, Ishita Bhansali, Q Vera Liao, Khai Nhut Truong, "FarmChat: A Conversational Agent to Answer Farmer Queries", *Proceedings of the ACM on Interactive, Mobile, Wearable and Ubiquitous Technologies,* December 2018 Article No. 170, https://doi.org/10.1145/3287048

[7] Sandeep Kumar Mohapatra; Anamika Upadhyay, "Using TF-IDF on Kisan Call Centre Dataset for Obtaining Query Answers," *2018 International Conference on Communication, Computing and Internet of Things,* DOI: 10.1109/IC3IoT.2018.8668134

[8] GUSTAVO MARQUES MOSTAÇO, ÍCARO RAMIRES COSTA DE SOUZ, LEONARDO BARRETO CAMPOS, CARLOS EDUARDO CUGNASCA, "AgronomoBot: a smart answering Chatbot applied to agricultural sensor networks", June 2018, *Conference: 14th International Conference on Precision Agriculture At: Montreal, Quebec, Canada,*

[9] Erik Cambria, Bebo White, "Jumping NLP Curves: A Review of Natural Language Processing Research", *May 2014 IEEE Computational Intelligence Magazine,* DOI: 10.1109/MCI.2014.2307227

[10] Giha Lee, Sungho Jung, "Application of Long Short-Term Memory (LSTM) Neural Network for Flood Forecasting," *July 2019 ResearchGate,* DOI: 10.3390/w11071387

[11] Dzmitry Bahdanau, Kyunghyun Cho, Yoshua Bengio, "Neural Machine Translation by Jointly Learning to Align and Translate", *Accepted at ICLR 2015 as oral presentation,* arXiv:1409.0473 [cs.CL]

[12] Minh-Thang Luong Hieu Pham Christopher D. Manning, "Effective Approaches to Attention-based Neural Machine Translation,", *2015 Conference on Empirical Methods in Natural Language Processing,* pages 1412–1421.

[13] Alex Sherstinsky, "Fundamentals of Recurrent Neural Network (RNN) and Long Short-Term Memory (LSTM) Network", *Elsevier "Physica D: Nonlinear Phenomena" journal,* Volume 404, March 2020: Special Issue on Machine Learning and Dynamical Systems

[14] Klaus Greff, Rupesh Kumar Srivastava, Jan Koutník, Bas R. Steunebrink, Jürgen Schmidhuber, "LSTM: A Search Space Odyssey", October 2017, *IEEE Transactions on Neural Networks and Learning Systems,* Volume: 28, Issue: 10, Oct. 2017

[15] Amir Jalilifard, Vinicius Caridá, Alex Mansano, Rogers Cristo, "Semantic Sensitive TF-IDF to Determine Word Relevance in Documents", *Information Retrieval (cs.IR); Computation and Language (cs.CL); Machine Learning (cs.LG); Machine Learning (stat.ML),* arXiv:2001.09896

[16] Akshay Nautiyala, Deepa Gupta, "KCC QA Latent Semantic Representation using Deep Learning & Hierarchical Semantic cluster Inferential Framework", *Third International Conference on Computing and Network Communications (CoCoNet'19),* Procedia Computer Science 171 (2020) 263–27

[17] Shely Koshy, N. Kishore Kumar, "Attitude of Farmers towards Kisan Call Centres", December 2012, *Journal of Extension Education,* 28(4):5753

[18] S. Kavitha, Nallusamy Anandaraja, "Kisan Call Centre Services to the Farming Community: An Analysis", September 2018, *Journal of Extension Education,* 29(3):5910

[19] Vasant P. Gandhi, Nicky Johnson, "Decision-Oriented Information Systems for Farmers: A Study of Kisan Call Centres (KCC), Kisan Knowledge Management System (KKMS), Farmers Portal, and M-Kisan Portal in Gujarat", *Centre for Management in Agriculture (CMA) Indian Institute of Management Ahmedabad (IIMA), Supported by Ministry of Agriculture and Farmers Welfare Government of India,* March 2018,

[20] B.R. Sharma, Pratap Singh and Amresh Sharma, "Role of Kisan Call Centres in Hill Agriculture", *Ind. Jn. of Agri. Econ. Vol.66, No.3, July-Sept. 2011.*

[21] Tom Bocklisch, Joey Faulkner, Nick Pawlowski, Alan Nichol, "RASA: Open Source Language Understanding and Dialogue Management", *NIPS Workshop on Conversational AI*